\documentclass[twocolumn,superscriptaddress,aps,pra]{revtex4-2}
\usepackage{graphicx}
\usepackage{dcolumn}
\usepackage{bm}
\usepackage[utf8]{inputenc}
\usepackage[T1]{fontenc}
\usepackage{mathptmx} 
\usepackage{bigints}

\begin{document}

\title{Peculiar effect of sample size in layered superconductors}

\author{Kesharpu Kaushal Kumar}
\email[Corresponding author: ]{kesharpu@theor.jinr.ru}
\affiliation{Joint Institute for Nuclear Research 141980 Dubna Russia}

\author{Kochev Vladislav Dmitrievich}
\affiliation{National University of Science and Technology <<MISiS>> 119049 Moscow Russia}

\author{Grigoriev Pavel Dmitrievich}
\affiliation{L. D. Landau Institute for Theoretical Physics RAS 142432 Chernogolovka Russia}
\affiliation{National University of Science and Technology <<MISiS>> 119049 Moscow Russia}

\date{\today}

\begin{abstract}
 We discuss an analytical model to calculate the superconducting volume ratio. Apart from this, our model can also predict the shape of embedded superconducting domains. We applied our model to calculate the superconducting volume ratios and shape of domains in (TMTSF)$_2$PF$_6$, (TMTSF)$_2$ClO$_4$, YBa$_2$Cu$_4$O$_8$, $\beta$-(BEDT)TTF$_2$I$_3$ and FeSe. Usually in layered superconductors resistivity drops anisotropically. Our analysis also explains that, this behaviour is due to flat or needle shape of the superconducting samples. 
\end{abstract}

\maketitle

\section{Introduction}

Some of the widely studied superconductors---cuprates\cite{shen-CuprateHighTcSuperconductors-2008} , iron based superconductors (IrSC) \cite{si-HightemperatureSuperconductivityIron-2016} and organic superconductors (OrSC) \cite{brown-OrganicSuperconductorsBechgaard-2015}---have layered structure. Usually, in these materials superconducting (SC) phase coexists with other electronic phases. In spatial coordinates this coexistence results in formation of SC domains inside other electronic phase. This behaviour is known as unconventional superconductivity \cite{stewart-UnconventionalSuperconductivity-2017}. Previously, resistivity experiments in cuprates and IrSC have shown anisotropic drop in resistivity. It was observed that, in identical temperature range, drop in resistivity along lowest conducting z-axis is higher compared to other two axis. In quasi one dimensional OrSC, (TMTSF)$_2$PF$_6$, anisotropic resistivity drop results in anisotropic SC transition temperature T$_c$. Although these facts have long been known, satisfactory theoretical explanation was not available.

Other problem in layered superconductors was absence of theoretical method to calculate the SC volume ratio in these materials. Previously Yonezawa \textit{et al.} \cite{yonezawa-CrossoverImpuritycontrolledGranular-2018}, Vuletic \textit{et al.} \cite{vuletic-CoexistenceSuperconductivitySpin-2002}, Pouget \textit{et al.} \cite{pouget-XrayEvidenceStructural-1983} and Kagoshima \textit{et al.} \cite{kagoshima-QuenchingEffectAnion-1983} tried to find the SC volume ratio in OrSC. However, there were some serious drawbacks in their assumptions. To calculate SC volume ratio Yonezawa \textit{et al.} assumed that, at SC transition temperature T$_{c}$ the whole sample becomes SC. However, due to granular superconductivity---presence of SC domains in background materials---it is not necessary. Because in this case SC domains can percolate to open SC channel. Also, they used the effective medium theory for their calculation. For successful application of effective medium theory: (I) SC domains should be spherical, (II) conductivity of background phase should be isotropic and (III) concentratio of SC phase should be small. For OrSC first two conditions are never satisfied. Hence, effective medium theory is not applicable in the proposed form.

Vuletic \textit{et al.} used a fractal model to calculate superconducting volume ratio. They assumed that, OrSC samples consist of alternating channels of spin density wave (SDW) and metallic phase. With decreasing temperature the cross-section of metallic channel decreases. When temperature T<T$_c$, metallic channels become superconducting. However, experiments have shown the presence of superconducting phase above T>T$_c$. Vuletic \textit{et al.} did not take into account this fact.  Similar drawbacks can be seen in the work of Pouget \textit{et al.} and Kagoshima \textit{et al.}

In series of work\cite{sinchenko-GossamerHightemperatureBulk-2017,grigoriev-AnisotropicEffectAppearing-2017,seidov-ConductivityAnisotropicInhomogeneous-2018,mogilyuk-ExcessConductivityAnisotropic-2019,kesharpu-SelfconsistentMaxwellApproximations-2019,kochev-AnisotropicZeroresistanceOnset-2021,kesharpu-EvolutionShapeVolume-2021}, we proposed a logically sound analytical model to calculate the superconducting volume ratio in the layered superconductors. The application of our model not only gave correct superconducting volume ratio, but also, explained: (I) the anisotropic resistivity drop in layered superconductors, (II) anistropic T$_c$ in (TMTSF)$_2$PF$_6$ and (III) increase in T$_c$ in IrSC due to decrease in width of the samples \cite{mogilyuk-ExcessConductivityAnisotropic-2019}, (IV) gave the shapes of SC domains in the samples and (V) predicted the onset of superconductivity in cuprates at T=250 K \cite{seidov-ConductivityAnisotropicInhomogeneous-2018}.

In the next section we briefly explained the theory of our model, then we discuss the results of application of our model.

\section{Theory}

\subsection{Space dilation and Maxwell-Garnett approximation}

In our approach, conductivity will be used to predict superconducting volume ratio $\phi$ using Maxwell-Garnett approximation (MGA) \cite{markel-IntroductionMaxwellGarnett-2016,markel-MaxwellGarnettApproximation-2016}. All the above mentioned materials are highly anisotropic. However, MGA can only be applied to isotropic medium. Hence, in first part of this section, a mapping procedure to convert anisotropic media to isotropic media is presented. Mapping is done obeying two conditions: (i) conductivity in mapped space along all 3 axes will be same; (ii) stationary current equation will be satisfied both in real space and mapped space. Let $\sigma_{xx}^b, \sigma_{yy}^{b}, \sigma_{zz}^b$ are background resistivity in real space. Hence, stationary current equation in real space will be
\begin{equation}
  \label{eq:continuity-equation-real}
  - \nabla j = \sigma_{xx}^b \frac{\partial^2 V}{\partial x^2} + \sigma_{yy}^b \frac{\partial^2 V}{\partial y^2} + \sigma_{zz}^b \frac{\partial^2 V}{\partial z^2} = 0.
\end{equation}
Here, $V$ is applied potential; $j$ is electric current density. In mapped space let the isotropic conductivity be $\sigma^{*} = \sigma_{xx}^{b}$. Here, $\sigma_{xx}^b$ is taken as reference conductivity for mapped space. Hence, stationary current equation in mapped space will be
\begin{equation}
  \label{eq:continuity-equation-map}
  - \nabla j = \sigma^{*}\left(\frac{\partial^2 V}{\partial x'^2} + \frac{\partial^2 V}{\partial y'^2} + \frac{\partial^2 V}{\partial z'^2}\right) = 0.
\end{equation}
According to the mapping condition, both Eq. (\ref{eq:continuity-equation-real}) and (\ref{eq:continuity-equation-map}) should be equal. Hence, equations for space dillation are
\begin{equation}
  \label{eq:space-dilation}
  x' = x; \quad  \sqrt{\mu} \: y' = y; \quad  \sqrt{\eta} \: z' = z.
\end{equation}
Here $\mu$ and $\eta$ are dilation coefficients. It is represented as
\begin{equation}
  \label{eq:dilation-coeff}
  \mu = \frac{\sigma^b_{yy}}{\sigma_{xx}^b}; \quad \eta = \frac{\sigma_{zz}^b}{\sigma_{xx}^b}.
\end{equation}

Due to dilation of the space, embedded SC domains also deforms into ellipsoid in mapped space. If SC domains are spheres with diameter $a$ in real space, then in mapped space they transform to ellipsoids with semi-axes
\begin{equation}
  \label{eq:sphere-in-map}
   a_x^{*} = a; \quad a_y^{*} = \frac{a}{\sqrt{\mu}}; \quad a_z^{*} = \frac{a}{\sqrt{\eta}}.
 \end{equation}
 Here, $a_x^{*}$, $a_y^{*}$ and $a_z^{*}$ are semi-axes of ellipsoids in mapped space along x, y, z-axes respectively. If SC inclusions in real space are ellipsoids with semi-axes $a_x = a$, $a_y = \beta \: a$ and $a_z = \gamma \: a$, then, in mapped space it transforms to ellipsoids with semi-axes
 \begin{equation}
  \label{eq:ellipsoid-in-map}
   a_x^{*} = a; \quad a_y^{*} = \frac{\beta \: a}{\sqrt{\mu}}; \quad a_z^{*} = \frac{\gamma \: a}{\sqrt{\eta}}.
 \end{equation}

\begin{figure}[!tbh]
  \centering
  \includegraphics[width = 0.4\textwidth]{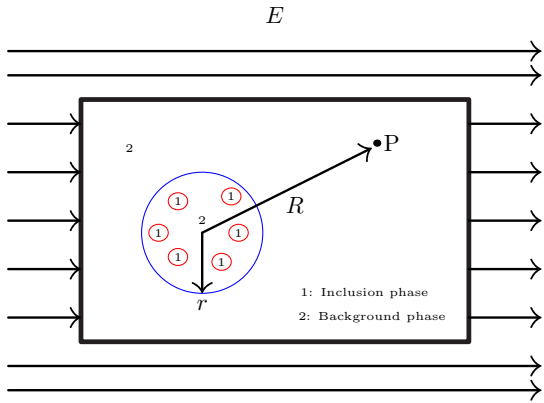}
  \caption{Schematic representation of Maxwell-Garnett approximation. The distance of point P from the center of sphere is very large compared to the size of the sphere, i.e. $r \ll R$. $E$ is the applied electric field.}
  \label{fig:Scheme-MGA}
\end{figure}
Before proceeding furgher, we explain the idea of MGA. To understand the basic idea of MGA refer Fig. \ref{fig:Scheme-MGA}. Here, we assume a sphere of mixed phase is embedded inside an infinite medium of background phase. Let, $E$ is the applied field; "1" denotes phase of the embedded domains, "2" denotes background phase, $R$ is the distance of some point $P$ from centre of the embedded sphere and $r$ is radius of the embedded sphere. The condition $R \gg r$ is satisfied. It is assumed that the size of the domains of phase "1" is small and distance between them is large. Hence, they don't interact with each other. This assumption is satisfied only if the volume ratio of phase "1" ($\phi_{1}$) is small. Hence, MGA is applicable only when $\phi_1 \ll 1$. The equation for effective conductivity is found from the condition that, electric field at point $P$ due to inclusions consisting of both phase "1" and "2" will be same due to inclusion with effective phase. According to this assumption, the condition
\begin{equation}
  \label{eq:MGA-Condition}
  (1 - \phi_1) (\sigma_e - \sigma_2) + \phi_1 \left[\frac{\sigma_2 \left(\sigma_e - \sigma_1\right)}{\sigma_2 + A_i (\sigma_1 - \sigma_2)}\right] = 0,
\end{equation}
is satisfied \cite{torquato-RandomHeterogeneousMaterials-2002}. Here, $\sigma_1$, $\sigma_2$ and $\sigma_e$ are conductivity of phase "1", "2" and effective phase respectively. $A_i$ is the depolarisation tensor along $i = x, \: y, \: z$ axes \cite{landau-ElectrodynamicsContinuousMedia-2008}. It occurs if the inclusions are ellipsoids. For an ellipsoid with semi-axes $a_1$, $a_2$ and $a_3$ the depolarisation tensor is represented as
\begin{equation}
  \label{eq:Depolarisation-Tensor}
  A_i = \prod\limits_{n=1}^3 a_n \bigintss dt/ 2 \left(t + a_i^2\right) \sqrt{\prod\limits_{n=1}^3 (t+a_n^2)}.
\end{equation}
For a spherical inclusion $A_x = A_y = A_z = 1/3$. Solution of Eq. (\ref{eq:MGA-Condition}) for $\sigma_e$ will give effective conductivity as
\begin{equation}
  \label{eq:MGA-Effective-Conductivity}
  \sigma_e = \sigma_2 \left\{ \frac{\left[A_i + (1 - A_i)\phi\right]\left(\sigma_1 - \sigma_2 \right) + \sigma_2}{A_i (1 - \phi_1) (\sigma_1 - \sigma_2) + \sigma_2}\right\}.
\end{equation}

\subsection{Temperature dependent shape of SC domains}
Magnetic susceptibility experiment in FeSe showed that, semi-axes ratio, $a_z/a_x$, of embedded SC domains increases with lowering of temperature \cite{grigoriev-AnisotropicEffectAppearing-2017}. As both FeSe and (TMTSF)$_2$PF$_6$ belong to UcSC \cite{stewart-UnconventionalSuperconductivity-2017}, we assume that, in (TMTSF)$_2$PF$_6$ also shape of embedded SC domains changes with temperature. We use resistivity experiment and Maxwell-Garnett approximation (MGA) to find shape of SC domains. The following assumptions were made for calculation: (I) the material consist of SC phase and background phase. Background phase is mixture of metallic and SDW phase. (II) SC domains are embedded inside background phase. These embedded SC domains have ellipsoidal shapes. (III) The distance between SC domains ($d$) is larger than their size ($l$), i.e. $d \gg l$. Hence, the interaction between SC domains can be neglected.

The procedure to find the shape of SC domains is as follows. First, an axis is chosen for calculation of superconducting volume ratio ($\phi$) from resistivity. Usually, lowest conducting axis is chosen, because, superconducting effect will be highest along lowest conducting axis. Using the resistivity along chosen axis and MGA approximation, $\phi$ is found by relation [See Ref. \cite{kesharpu-EvolutionShapeVolume-2021} and reference therein, for derivation of following formulas]
\begin{equation}
  \label{eq:phi}
  \phi = \frac{A_i \left[ 1 - \left( \sigma_{ii}^b/ \sigma_{ii}\right) \right]}{A_i \left[ 1 - \left( \sigma_{ii}^b/ \sigma_{ii}\right) \right] + \left( \sigma_{ii}^b/ \sigma_{ii}\right)}.
\end{equation}
Here, $\sigma_{ii}^b$ is background conductivity along i-th axis; it is the conductivity of the material without SC phase. $\sigma_{ii}$ is effective conductivity along i-th axis. It is the conductivity of the material with SC phase. A$_i$ is the depolarisation factor of ellipsoidal SC inclusions. It is defined as
\begin{equation}
  \label{eq:depolarization-factor}
  A_i = \sum\limits_{n=1}^3 a_n \bigint_0^{\infty} dt \bigg/ 2 \left( t + a_{i}^2\right) \sqrt{\sum\limits_{n=1}^3 (t + a_n^2)}.
\end{equation}
Using $\phi$ from Eq.~(\ref{eq:phi}), resistivity along other two axes is predicted. The predicted resistivity is
\begin{equation}
  \label{eq:conductivity}
  \sigma_{ii} = \sigma_{ii}^b \left[\frac{A_i + \left( 1 - A_i \right)\phi}{A_i (1 - \phi)}\right].
\end{equation}
The shape of SC domains is found when predicted resistivity in Eq.~(\ref{eq:conductivity}) coincides with experimental values, because, predicted resistivity depends $\phi$, which depends on shape dependent term A$_i$. 

\section{Results \& Discussion}

Using above method superconducting volume ratio was found in cuprates \cite{seidov-ConductivityAnisotropicInhomogeneous-2018}, IrSC \cite{grigoriev-AnisotropicEffectAppearing-2017,sinchenko-GossamerHightemperatureBulk-2017} and organic superconductors \cite{kesharpu-EvolutionShapeVolume-2021,kochev-AnisotropicZeroresistanceOnset-2021, seidov-ConductivityAnisotropicInhomogeneous-2018}. Apart from that, several other conclusion were made in these works.

In \textit{Seidov et al.} we predicted that in YBa$_2$Cu$_4$O$_8$ superconducting domains starts to appear at temperature T=250K \cite{seidov-ConductivityAnisotropicInhomogeneous-2018}. It is much higher than the superconducting transition temperature T$_c$. Similarly they showed that the semi-axes of the SC domains are proportional to the SC coherence lengths. We also found the superconducting volume ratio in organic superconductor $\beta$-(BEDT)-TTF$_2$I$_3$. These volume ratios were confirmed by magnetic susceptibility experiments.

In \textit{Sinchenko et al.} and \textit{Grigoreiv et al.} we predicted the volume ratio $\phi$ in FeSe \cite{sinchenko-GossamerHightemperatureBulk-2017,grigoriev-AnisotropicEffectAppearing-2017}. Using magnetic susceptibility the exact shapes of the superconducting domains were found in this work. It showed that the ratio of semi-axes $a_z/a_x \to 1$ as $T \to 0$. Similar behaviour was also seen in (TMTSF)$_2$Col$_4$ \cite{kesharpu-EvolutionShapeVolume-2021}. Both these cases infers that, SC islands do not originate from superconducting fluctuation.

\begin{widetext}
\begin{figure}[!tbh]
  \centering
  \includegraphics[width = 0.9\textwidth]{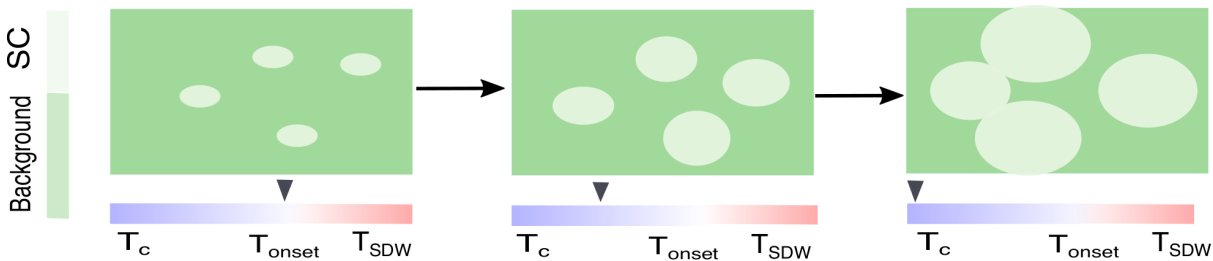}
  \caption{Schematic representation of evolution of embedded SC domains in background phase. SC domains starts to appear at T = T$_{onset}$. With decrease in temperature the shape and size of SC domains increases. At T=T$_c$ superconducting domains percolates to open superconducting channels.}
  \label{fig:Phase-Evolution}
\end{figure}
\end{widetext}

In \textit{Kochev et al.} we showed that, anisotropic superconducting transition temperature T$_c$ observed in the organic superconductor (TMTSF)$_2$PF$_6$ is due to shape of the samples. The explanation is as follows. Usually, synthesis of layered superconductors gives flat or needle shaped samples. The thickness of the samples is 2 order smaller than the length of the samples ($l_z \ll l_y \ll l_x$). It should be noted that z-axis is the lowest conducting axis. Hence, when the size of SC domains increase, the percolation probability of these domains will be highest along shortest z-axis. This result in percolation of SC domains along the lowest conducting z-axis. A schematic representation of this idea is shown in Fig. \ref{fig:Phase-Evolution}.
In (TMTSF)$_2$ClO$_4$ the superconducting transition is controlled by cooling rate. This behaviour is directly related to shape and size of SC domains. Hence, using our method we investigated the evolution of shape of the SC domains in \textit{Kesharpu et al.} We found that, at small disorders the SC domains are flatter along z-axis compared to higher disorder. Recent unpublished results has shown that, in (TMTSF)$_2$PF$_6$ pressure has the same effect.

\section{Conclusion}

We discussed an analytical model for calculation of superconducting volume ratios. In this model we tried to alleviate the drawbacks of the previous model by \textit{Yonezawa et al.} and \textit{Vuletic et al.} Apart from that, our model can calculate the shape of the SC domains.

\section{Acknowledgement}
This article is partly supported by the Ministry of Science and Higher Education of the Russian Federation in the framework of Increase Competitiveness Program of MISiS by Russian Foundation for Basic Research (RFBR) Grant No. 21-52-12027 and by the “Basis” Foundation for the development of theoretical physics and mathematics. V.D.K. acknowledges the MISiS Project No. K2-2020-001, and K.K.K. acknowledges the MISiS support project for young research engineers and RFBR Grants No. 19-32-90241. P.D.G. acknowledges the State Assignment No. 0033-2019-0001 and RFBR Grants No. 19-02-01000 and No. 21-52-12043.

\bibliographystyle{apsrev4-2}
\bibliography{zoterolibrary.bib}

\newcommand{\noopsort}[1]{}

\end{document}